\documentclass[twocolumn,showpacs,preprintnumbers,amsmath,amssymb]{revtex4}

\usepackage{graphicx}
\usepackage{dcolumn}
\usepackage{bm}
\begin{document} 

\title{From cluster to solid - the variational cluster approximation
applied to NiO}
\author{R. Eder}
\affiliation{Forschungszentrum Karlsruhe, 
Institut f\"ur Festk\"orperphysik, 76021 Karlsruhe, Germany}
\date{\today}

\begin{abstract}
The variational cluster approximation is applied to the calculation of
the single particle spectral function of NiO.
Trial self energies and the numerical value of the Luttinger-Ward functional
are obtained by exact diagonalization of $NiO_6$-clusters and
the single particle parameters of the clusters serve as variational 
parameters to obtain a stationary point of the grand potential of
the lattice system. Good agreement with experiment is obtained.
\end{abstract} 

\pacs{72.80.Ga,71.27.+a,79.60.-i}

\maketitle

\begin{figure}
\includegraphics[width=\columnwidth]{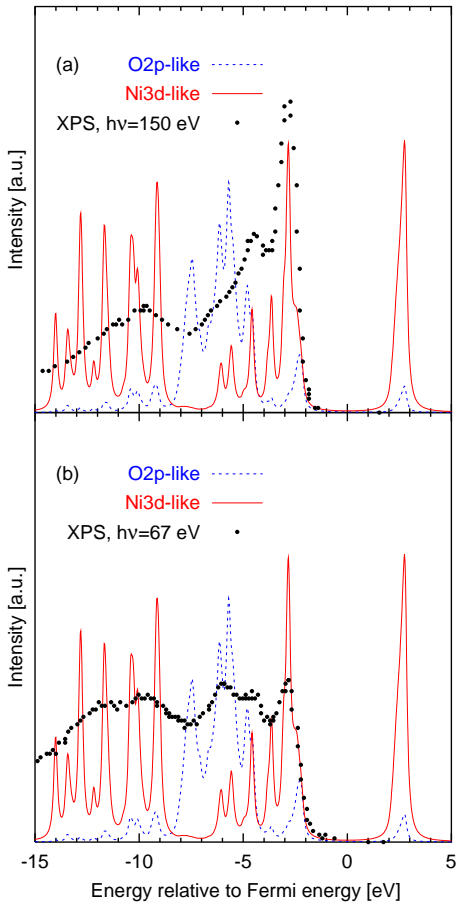}
\caption{\label{fig1} Single particle spectral densities
obtained by VCA compared to valence band photoemission data (XPS).}
\end{figure}

\begin{figure}
\includegraphics[width=\columnwidth]{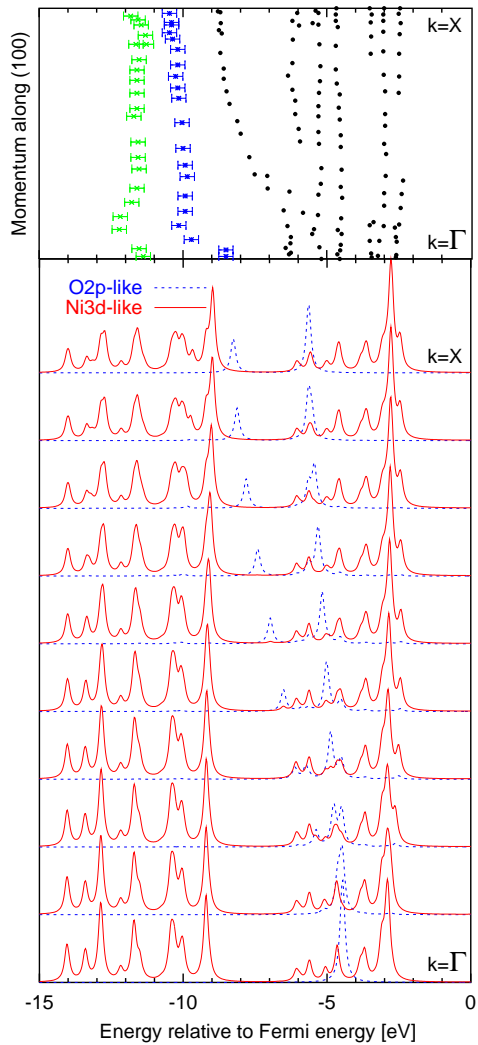}
\caption{\label{fig2} Top: Dispersion of experimental bands
measured by ARPES\cite{Shen_long}. Bottom: $\bm{k}$-dependent
spectral function for momenta along $\Gamma-X$. Lorentzian broadening
$0.1\;eV$, $d$-like weight is multiplied by factor of $5$.}
\end{figure}

The theoretical description of compounds containing partially
filled $3d$, $4f$ or $5f$ shells is a much-studied problem in
solid-state theory. Due to the small spatial extent of these shells
the Coulomb repulsion between conduction electrons in these compounds
becomes unusually strong 
and approximations which rely on a mapping of the
physical electron system onto one of fictious free particles in
a suitably constructed effective potential cannot even qualitatively
describe the resulting state. Starting with the work of 
Hubbard\cite{Hubbard} a variety of theoretical methods have been invented to
deal with this 
problem\cite{SvaGu,Czyzyk,Bala,Manghi,Ary,Iga,Kunesetal,Massi}.
Major progress towards a quantitative
description of $3d$ transition metal (TM) oxides has been made by the 
cluster method initiated
by Fujimori and Minami\cite{FujimoriMinami,Elp}. This takes the opposite
point of view as compared to band theory, namely to abandon
translational invariance  and instead treat exactly - by means of
atomic multiplet theory\cite{Slater,Griffith} - the
Coulomb interaction in the $3d$-shell of a TM-ion in an octahedral `cage'
of nearest-neighbor oxygen atoms. 
Recently ideas have been put forward to broaden the spectra of
finite clusters into bands\cite{Senechaletal,Maier,Aichhorn}.
In particular, building on field-theoretical work of
Luttinger and Ward\cite{LuttingerWard} who showed that the
grand canonical potential $\Omega$ of an interacting Fermion system
is stationary with respect to variations of the electronic self-energy
$\Sigma$, Potthoff has recently proposed\cite{PotthoffI} the variational
cluster approximation (VCA) where trial self energies are generated
numerically by solving finite clusters and used
in a variational scheme for $\Omega$. So far the VCA has been applied 
mainly to simplified systems such as
the single-band Hubbard-model\cite{PotthoffI,Dahnken} but the success of the 
cluster method for TM-oxides suggests to apply the VCA also to a realistic
model for TM-oxides thereby using the octahedral clusters
discussed above to generate self-energies.
Here we outline such a calculation for the frequently studied
compound NiO. Using clusters containing just a single TM-ion
implies that the self-energy is site-diagonal, i.e.
${\bm k}$-independent. The
corresponding approximation thus is similar to the dynamical
mean-field calculations which have recently been applied to
a variety of compounds\cite{DMFTs}.\\
We start by defining the Hamiltonian which describes the NiO lattice
and denote by
$d_{i,\alpha,\sigma}^\dagger$ an operator
which creates a spin-$\sigma$ electron in the
$d$-orbital $\alpha\in \{ xy, xz, yz,\dots\}$ on metal site $i$
and $p_{j,\lambda,\sigma}$ annihilates an electron
in $p$-orbital $\lambda\in \{ x,y,z\}$ on oxygen site $j$.
Taking the O$2p$-level energy $\epsilon_p$ as the zero of energy
the single-particle terms read
\begin{eqnarray}
H_{0} &=&\sum_{i,\alpha,j,\lambda} \sum_\sigma\;
(t_{i,\alpha}^{j,\lambda}\;  d_{i,\alpha,\sigma}^\dagger\;
p_{j,\lambda,\sigma}^{} + H.c.) \nonumber \\
&&\;\;\;\;\;\;\;\;\;+\;
\sum_{i,\alpha,\sigma} (\epsilon_d + \epsilon_\alpha)
d_{i,\alpha,\sigma}^\dagger\;
d_{i,\alpha,\sigma}^{}.
\label{hybrid}
\end{eqnarray}
The Hamiltonian also contains
terms which describe hybridization between
next-nearest neighbors, i.e. O$2p$-O$2p$ and  Ni$3d$-Ni$3d$.
The numerical values of the different hopping integrals
have been obtained by an LCAO-fit to a paramagnetic LDA
band structure of NiO\cite{nio}. They are quite similar to those
used by Fujimori and Minami\cite{FujimoriMinami}
and van Elp {\em et al.}\cite{Elp}.
Also included - via the $\epsilon_\alpha$ -
is a CEF-splitting of $10Dq=0.7eV$\cite{FujimoriMinami,Elp}.\\
The Coulomb interaction within the $d$-shell reads
\begin{equation}
H_{1}=\sum_{\lambda_1,\lambda_2,\lambda_3,\lambda_4}
V_{\lambda_1,\lambda_2}^{\lambda_3,\lambda_4}\;\;
 d_{\lambda_1}^\dagger  d_{\lambda_2}^\dagger
d_{\lambda_3}^{} d_{\lambda_4}^{},
\label{inter}
 \end{equation}
where we have suppressed the site label $i$ and $\lambda=(\alpha,\sigma)$.
The matrix elements
$V_{\lambda_1,\lambda_2}^{\lambda_3,\lambda_4}$
can be expressed\cite{Slater,Griffith} in terms of
the $3$ Racah-parameters, $A$, $B$ and $C$.
To avoid an `implicit' interaction between electrons
these are assumed to be independent of the $d$-shell occupation.
$B$ and $C$ can be estimated from atomic Hartree-Fock wave
functions and the values used here, $B=0.13eV$ and $C=0.6eV$ are standard 
ones\cite{FujimoriMinami,Elp}.
The parameter $A$ is reduced substantialy by solid state screening
and is usually adjusted to match experiment - here we do the same and choose 
$A=9eV$.
The `Hubbard $U$'$=E(d^{9}) + E(d^{7}) - 2 E(d^8)$ - calculated from 
ground state
energies of free ions without CEF-splitting - is\cite{Griffith}
 $U=A+B=9.1eV$. 
Previous estimates range from $U=6.7 eV$ - by fit of
cluster spectra to experiment\cite{FujimoriMinami,Elp} - to
$U=8eV$ by density functional calculations\cite{Anisimov}.
The present value thus is somewhat large which
will be discussed below in more detail.
Next, we choose - again by adjusting to experiment -
$\epsilon_d-\epsilon_p=-68.5\;eV$ so that 
the charge transfer energy
$\Delta=E(d^{9}\underline{L})-E(d^8)$ is\cite{Griffith}
$\Delta=\epsilon_d-\epsilon_p +8A-6B+7C=6.9 eV$.
These values are consistent with
the notion that NiO is a charge transfer insluator but
close to the intermediate regime of the 
Zaanen-Sawatzky-Allen diagram\cite{ZSA}. 
Finally, any Coulomb interaction between electrons which are
not in the same $Ni3d$-shell is neglected.\\
The VCA is based on an expression for the  grand potential $\Omega$
of an interacting many-Fermion system due to
Luttinger and Ward\cite{LuttingerWard}. In a multi-band system
where the Green's function ${\bf G}({\bf k},\omega)$,
the noninteracting kinetic energy ${\bf t}({\bf k})$
and the self-energy ${\bf \Sigma}({\bf k},\omega)$
for given energy $\omega$ and momentum ${\bf k}$ are matrices of
dimension $2n\times 2n$, with $n$ the number of 
orbitals/unit cell, it reads\cite{Luttingertheorem}
\begin{eqnarray}
\Omega &=& -\frac{1}{\beta}\sum_{{\bf k},\nu}e^{\omega_\nu 0^+}
\ln \;det\left(-{\bf G}^{-1}({\bf k},\omega_\nu)\right)+ 
F[{\bf \Sigma}]
\label{ydef}
\end{eqnarray}
where $\omega_\nu=(2l+1)\pi/\beta$ are the 
Matsubara frequencies,
\begin{equation}
{\bf G}^{-1}({\bf k},\omega) =\omega + \mu - {\bf t}({\bf k})
- {\bf \Sigma}({\bf k},\omega)
\label{gdef}
\end{equation}
and the functional $F[{\bf \Sigma}]$ is the Legendre
transform\cite{PotthoffI} of the
Luttinger-Ward functional $\Phi[{\bf G}]$.
A definition of $\Phi[{\bf G}]$ in terms of Feynman diagrams
was given by Luttinger and Ward\cite{LuttingerWard}, a nonperturbative
derivation has recently been given by Potthoff\cite{Nonperturbative}.
$\Omega$ is stationary with respect to variations of the 
self-energy\cite{LuttingerWard}
\begin{equation}
\frac{\delta \Omega}{\delta \Sigma_{ij}({\bf k},\omega_\nu)} = 0
\label{stationary}
\end{equation}
but a prohibitive obstacle in exploiting (\ref{stationary})
in a variational scheme for
${\bf \Sigma}$ is the evaluation of $F[{\bf \Sigma}]$ 
for a given `trial ${\bf \Sigma}$'. 
Potthoff has suggested\cite{PotthoffI} to 
restrict the domain of ${\bf \Sigma}$ to
`cluster representable' ones, i.e. exact self-energies of
finite clusters, for which 
$F[{\bf \Sigma}]$  can be determined numerically from (\ref{ydef}).
The key observation\cite{PotthoffI} is that $\Phi[{\bf G}]$ and hence its
Legendre transform $F[{\bf \Sigma}]$
have no explicit dependence on the single-particle terms
of $H$ whence $F[{\bf \Sigma}]$
is the same functional of ${\bf \Sigma}$ for any two systems
with the {\em same interaction part} of the Hamiltonian.
Under the assumption that only interaction lines connecting
orbitals in the same d-shell are relevant we can therefore
use a numerically soluble system of disconnected finite clusters - 
the so-called reference system - to generate trial self-energies
${\bf \Sigma}(\omega)$ together with their exact $F[{\bf \Sigma}]$.
More precisely we choose a reference system where each $Ni3d$ 
orbital $d_\alpha$ 
is coupled to one `ligand' orbital $L_\alpha$ with these ligands in turn
decoupled from each other and the interaction
within the $d$-shell given by (\ref{inter}). The reference system thus is
equivalent to an array of non-overlapping identical
NiO$_6$ clusters where each ligand $L_\alpha$
corresponds to the unique linear combination of $O2p$
orbitals on the six nearest $O$ neighbors of a given Ni atom which hybridizes
with the $Ni3d_\alpha$ orbital. After numerical diagonalization 
of the cluster Hamiltonian for all possible electron numbers
we obtain the grand potential $\tilde{\Omega}$
and Green's function $\tilde{{\bf G}}(\omega)$ of
the cluster whence equations (\ref{gdef}) and (\ref{ydef}) give 
${\bf \Sigma}(\omega)$ and $F[{\bf \Sigma}]$.
Next we insert the ${\bf \Sigma}(\omega)$ and $F[{\bf \Sigma}]$ so
obtained into equations (\ref{ydef}) and (\ref{gdef})
{\em for the lattice system} and obtain an approximate
${\bf G}(\bf{k},\omega)$ and $\Omega$ for the infinite system.
Variation of ${\bf \Sigma}(\omega)$ is performed by varying the
single-electron parameters - such as hybridization integrals or
site-energies - of the reference system and the
best approximation to ${\bf \Sigma}(\omega)$
is obtained by demanding that $\Omega$ be stationary
with respect to such variations.  
We write the single-particle Hamiltonian for a NiL$_5$ cluster as
\begin{eqnarray}
H_{single} &=& \sum_{\alpha,\sigma}\;V(\alpha)\left(\;d_{\alpha,\sigma}^{\dagger}
L_{\alpha,\sigma}^{} + H.c.\;\right) \nonumber \\
&& + \sum_{\alpha,\sigma}
\left(E({\alpha})\;d_{\alpha,\sigma}^{\dagger}
d_{\alpha,\sigma}^{} + e({\alpha}) \;L_{\alpha,\sigma}^{\dagger}
L_{\alpha,\sigma}^{}\right)
\label{heff}
\end{eqnarray}
and have the following varational parameters:\\
1) The hopping integrals $V(\alpha)$. Since the ground state of $d^8$
in cubic symmetry has the configuration $t_{2g}^6 e_g^2$,
$V(t_{2g})$ connects mainly completely occupied orbitals.
To simplify the problem we therefore
discard the three $t_{2g}$-like ligands alltogether and
write the remaining $V(e_g)=\lambda \tilde{t}$
where $\tilde{t}=\sqrt{3}(pd\sigma)$ is the hopping integral
in the cluster calculation for a NiO$_6$-cluster\cite{FujimoriMinami}.\\
2) The site energy of the $e_g$-like ligands $e(e_g)$.\\
3)  The site energies $E({\alpha})$ of the $e_g$ and $t_{2g}$-like d-orbitals.
By numerical scan a set of these $4$ parameters 
where $\Omega$ is stationary can be found for each temperature $T$.
For temperatures between $100-1000 K$ the expression $\Omega(T) =
\Omega_0 - k_B T \log(3)$ - exptected for a gapped, paramagnetic
spin-1 system - gives an excellent fit to the calculated $\Omega(T)$.
The constant $\Omega_0$ thereby is the ground state expectation
value $\langle H-\mu N\rangle$. Since the VCA also gives the lattice Green's
function $\bf{G}(\bf{k},\omega)$, 
$\langle H-\mu N \rangle$ can alternatively be computed from its $0^{th}$
and $1^{st}$ moments.
There is no reason why these two results should agree - the formalism of
the VCA would not necessitate this. Still
the two values agree quite well - $-418.649\;eV$/unit cell
versus $-418.684\;eV$/unit cell - indicating that the
VCA is a quite `intrisically consistent' approximation.
Next, we proceed to a comparison of  $\bf{G}(\bf{k},\omega)$ to
experiment.
Figure \ref{fig1} compares $\bm{k}$-integrated
spectral densities calculated at room temperature $(300$ Kelvin) to
angle integrated valence band photoemission spectra
taken by Oh {\em et al.}\cite{Ohetal} 
at two different photon energies.
At $h\nu=150eV$ the experimental spectrum essentially resembles
the $d$-like spectral density, whereas 
at $h\nu=67eV $ the states at the valence band top
are anti-resonantly supressed - whence $O2p$-derived features become 
more clearly visible - whereas the 
`satellite' at $-10eV$ is resonantly enhanced. 
Figure \ref{fig2} compares the $\bm{k}$-resolved spectral function
for momenta along $(100)$ $(\Gamma \rightarrow X)$ 
to the experimental band dispersion by Shen {\em et al.}\cite{Shen_long}.\\
The spectral density has gap of approximately $4eV$ around
the chemical potential. This is consistent
with experiment\cite{SawatzkyAllen} but has of course been achieved by
the choice of $A$ and $\Delta$.
At the top of the photoemission spectrum $E<0$
there is a high-intensity band complex
at binding energies between $\approx -4\;eV $
and $\approx -2\;eV $, which was shown to
consist of several sub-peaks by Shen {\em et al.}\cite{Shen_long}.
These authors did not actually resolve the dispersion of the
individual sub-peaks
although the data seem to indicate a weak overall `upward'
dispersion as one moves $\Gamma \rightarrow X$
which would be consistent with theory. It also has to be kept in mind
that the calculation has been performed for the paramagnetic phase
whereas the experiment was done below the N\'eel temperature
and thus in the antiferromagnetically ordered phase.
This may have an impact on the dispersive features.
Proceeding to more negative binding energy both the experimental
band structure and the theoretical spectra show a gap of
$\approx 1eV$ and then a group of dispersionless bands
between $-4.5eV$ and $-6.5eV$.
In the angle-integrated spectrum, Figure \ref{fig1}a,
the topmost of these bands produces the shoulder at 
$-4.5 eV $. The spectrum
in Figure \ref{fig1}b shows a peak at $-6eV$ which originates from
states with essentially pure $O2p$ character\cite{Ohetal}.
The corresponding peak in the theoretical spectrum originates from a
saddle point singularity of the upper $O2p$ derived band at the $X$-point.
Comparison of the angle integrated spectrum, the experimental
dispersion and an LDA band structure makes this a plausible
explanation. The sole strongly dispersive feature
in the spectrum, namely an $O2p$-derived band at binding
energies between $-6eV\rightarrow -9eV$ is again well reproduced by theory.
Finally the `satellite' at binding
energies $-9eV\rightarrow -15eV$ consists of at least two sub-peaks
as can be seen in Figure \ref{fig1}b and also in the ARPES data. 
Theory predicts several sub-peaks in the satellite but these may
not have been resolved in experiment due to the strong broadening
of the satellite. 
By and large we may say that there is essentially a one-to-one
correspondence between theory and the measured bands 
(the $\bf{k}$-integrated spectra also agree roughly
with a recent LDA+DMFT(QMC)-calculation\cite{Kunesetal}).
Finally we mention the values of the $d$-shell occupation $n_d=8.16$
and the expectation value of the $d-p$ hybridization 
$\langle H_{pd}\rangle=-2.30\; eV$/unit cell
(at $T=300\;K$ with negligible $T$-dependence). 
To put these in perspective we note that perturbation theory for
a NiO$_6$ cluster with unrenormalized parameters gives
$n_d=8+2(\tilde{t}/\Delta)^2=8.21$ and 
$\langle H_{pd}\rangle=-4\tilde{t}^2/\Delta=-2.93eV$.\\
An issue that requires more detailed discussion is the spectral weight
of the satellite:  the fraction of
$d$-like weight below the`gap' around $-7eV$ is $w=61\%$.
A recent LDA+DMFT calculation\cite{Kunesetal} gave $w=45\%$ with $U=8eV$.
Van Elp {\em et al.}\cite{Elp} estimated $w=30\%$ with $U=6.7eV$.
As shown in Ref.\cite{Elp} 
the spectral weight of the satellite is directly related to
the choice of $U$ - larger $U$ produces a more intense satellite.
The relatively large value of $U=9.1eV$ in the present
work was necessary to get the satellite below the bottom of the
$O2p$-band - as suggested by ARPES\cite{Shen_long}.
This raises the question why we need a larger value of $U$ as compared
to the $6.7eV$ in the cluster calculations\cite{FujimoriMinami,Elp}.
It turns out that the reason is the downward renormalization of the
hopping integral $V(e_{g})$ in (\ref{heff}). 
The value of $\lambda$ at the stationary point
is $0.812$ (at $300$K with negligible $T$-dependence)
and even such a small reduction leads to an appreciable upward shift
of the satellite due to reduced level repulsion between satellite
and valence band top (see e.g. Figure 11 of Ref.\cite{Elp}) - which
must be compensated by a larger $U$.
On the other hand there is a simple physical argument
for the value of $\lambda<1$: in a  NiO$_6$ cluster, the
mixing strength between (say) a $Ni3d_{x^2-y^2}$ orbital and the
bonding combination of $O2p_x$ orbitals on the two nearest neighbors
in $x$-direction is $\tilde{t}/\sqrt{2}$. In a NiO lattice
the matrix element between the Bloch states of momentum $\bf{k}$
formed from the $Ni3d_{x^2-y^2}$ orbital and the $O2p_x$-orbital is
$i\tilde{t}\sin(k_x a)$ with $a$ the Ni-O distance.
The Brillouin zone average of $|\sin(k_x a)|$ is $0.63=0.89/\sqrt{2}$
so that the value of $\lambda$ at least partly reflects the smaller
average $d-p$ hybridization in the NiO lattice as compared to the NiO$_6$
cluster. Anisimov {\em at al.}\cite{Czyzyk} gave a similar
argument to improve the agreement of their LDA+U calculation with 
the cluster calculations and in principle this
should occur in any approximation
where a $\bf{k}$ independent self-energy is determined in an
`impurity'-like calculation. More accurate knowledge about the
total weight and width of the satellite probably
will be needed to decide which is the more correct description.
Finally it should be noted that the values of $U$ and $\Delta$ determine
mainly the energies of the upper Hubbard band and satellite
relative to the valence band top - the band
structure above $-7eV$ is influenced hardly at all
by these parameters.\\
In summary: the variational cluster approximation due to Potthoff
allows to combine the powerful
cluster or CI method for transition metal compounds
with the field-theoretical work of Luttinger and Ward to implement a
variational scheme for the electronic self-energy and construct
a band structure method for  strongly correlated electron
compounds. Both, a realistic band structure
and the full atomic multiplet interaction can be incorporated into
the Hamiltonian without problems, the system can be studied
at arbitrarily low temperatures and the Green's function be obtained
with arbitrary energy resolution. 
The results are quite encouraging in that there is
an  essentially one-to-one correspondence between calculated
Green's function and electron spectroscopies whereby
a comparison of the fine structure of the
broad band complex at the valence band top to experimental data
of higher resolution would be desirable. 
The good agreement also suggests that the band structure
of NiO is `Coulomb generated' in that the atomic multiplet
structure survives with minor modifications and gets
broadened into weakly dispersive bands. All in all the VCA
appears to be a promising tool for the study of transition metal
compounds.\\
The author would like to thank M. Potthoff for instructive discussions.

\end{document}